\documentclass[journal,final]{article}

\usepackage{amsmath,epsfig,amsfonts,dsfont}

\newcommand{\diag}{\textrm{diag}}

\newcommand{\A}{\mathbf{A}}

\newcommand{\C}{\mathbf{C}}
\newcommand{\D}{\mathbf{D}}
\newcommand{\F}{\mathbf{F}}
\newcommand{\G}{\mathbf{G}}

\newcommand{\X}{\mathbf{X}}
\newcommand{\Y}{\mathbf{Y}}
\newcommand{\U}{\mathbf{U}}
\newcommand{\V}{\mathbf{V}}
\newcommand{\x}{\mathbf{x}}

\newcommand{\vet}{\mathbf{v}}
\newcommand{\y}{\mathbf{y}}
\newcommand{\z}{\mathbf{z}}
\newcommand{\uvet}{\mathbf{u}}
\newcommand{\f}{\mathbf{f}}
\newcommand{\g}{\mathbf{g}}
\newcommand{\bgamma}{\boldsymbol{\gamma}}
\newcommand{\id}{\mathds{1}}
\newcommand{\R}{\mathds{R}}
\newtheorem{theorem}{Theorem}
\newtheorem{corollary}{Corollary}

\title{Phase space decomposition for phase noise and synchronization analysis of planar nonlinear oscillators}

\author{Michele Bonnin\thanks{The authors are with the Department of
		Electronics and Telecommunications, Politecnico di Torino, Turin,
		Italy (e-mail: michele.bonnin@polito.it).}, Fernando Corinto,  Marco Gilli}

\begin{document}

\maketitle

\begin{abstract}
Synchronization phenomena, frequency shift and phase noise are
often limiting key factors in the performances of oscillators. The
perturbation projection method allows to characterize how the
oscillator's output is modified by these disturbances. In this
brief we discuss the appropriate decomposition of perturbations
for synchronization and phase noise analysis of planar nonlinear
oscillators. We derive analytical formulas for the vectors
spanning the directions along which the perturbations have to be
projected. We also discuss the implications of this decomposition
in control theory and to what extent a simple orthogonal
projection is correct.
\end{abstract}

\section{Introduction}\label{intro}

Oscillators are ubiquitous in modern electronic and optical
devices. For instance, in digital systems they are responsible to
give a reference signal to synchronize operations. In radio
communication systems they are used for frequency and amplitude
modulations to convey information.

An ideal oscillator would exhibit a perfectly localized spectrum
at the desired frequency. However, the output of actual
oscillators is always corrupted by external disturbances, internal
noise source, thermal noise, and interaction with other
oscillators. As a consequence the power spectra of practical
oscillators exhibit both a shift of its peaks and linewidth
broadening. These phenomena, often referred to as frequency shift
and phase noise, are key performance limiting factors in
electronic systems. Characterizing how perturbations affect the
performances of oscillators is therefore a problem of both
theoretical and practical paramount importance, which has received
lot of attention since the seminal work \cite{lax67}.

The traditional approach is to linearize the system about the
unperturbed solution, assuming that the resultant deviation are
small \cite{okumura90,kartner89}. However it turns out that in
several situations the small deviation assumption becomes invalid
and the linearized perturbation analysis is inconsistent
\cite{kartner90,hajimiri98}. To overcome this limitation nonlinear
analysis methods, based on the decomposition of the perturbation
into orthogonal components has been proposed
\cite{kartner90,hajimiri98}. The method has later been improved
with the introduction of a non orthogonal decomposition, leading
to the so called \emph{perturbation projection method}
\cite{demir00,demir03}.

The perturbation projection method has been used to study
injection locking \cite{lai04,maffezzoni08}, pulling effects
\cite{maffezzoni09}, synchronization phenomena
\cite{maffezzoni10}, and power spectra of noisy oscillators
\cite{demir02,demir03,traversa11}. Different numerical schemes
have been proposed \cite{demir03,demir02,lai04,chow07,djurhuus09}
to find the proper decomposition of the perturbation, that has to
be projected onto two complementary subspaces. It turns out that
the bases of the two spaces are related to Floquet's eigenvectors,
and their determination requires several numerical integrations of
both linear and nonlinear differential problems, a potentially
burdensome task \cite{demir00,brambilla05,djurhuus09}.

This brief discusses the appropriate phase space decomposition for
planar nonlinear oscillators. The main contribution of the paper
is the derivation of analytical formulas for the vectors spanning
the subspaces along which the perturbations has to be decomposed.
Using a classical result on the integration of planar autonomous
differential equations we derive these formulas, that allow to
write a nonlinear differential equation for the time evolution of
the oscillator's phase deviation in a closed form. This equation
is the starting point to investigate synchronization phenomena,
frequency shift and phase noise in weakly perturbed oscillators.
We also discuss some implication of the decomposition in control
theory and to what extent the old orthogonal decomposition is
correct.


\section{Phase space decomposition of nonlinear oscillations}\label{decomposition}

We consider nonlinear oscillators subject to external
perturbations described by the ODE
\begin{equation}\label{sec2-eq1}
\dot{\x}(t) = \f(\x(t)) + \varepsilon \, \g(\x(t),t)
\end{equation}
where $\x : \R \mapsto \R^n$ is the state of the oscillator, $\f :
\R^n \mapsto \R^n$ describes the oscillator's internal dynamics,
$\g :\R^n \times \R \mapsto \R^n$ defines the perturbation and
$\varepsilon \ll 1$ measures the strength of the perturbation. In
absence of the perturbation, e.g. for $\varepsilon =0$, the
oscillator exhibits an asymptotically stable $T$--periodic limit
cycle $\bgamma$
\begin{equation}\label{sec2-eq2}
\left\{
\begin{array}{rcl}
\dot{\x}_0(t) & = & \f(\x_0(t))\\[1ex]
\x_0(t) & = & \x_0(t+T).
\end{array}\right.
\end{equation}

The ideal framework to investigate phase noise effects, phase
locking and synchronization phenomena in \eqref{sec2-eq1} are
\emph{phase models} \cite{demir00}. Phase models are based on the
idea to decompose the perturbation into two components, one
tangent and one transversal, but not perpendicular to, the
unperturbed limit cycle. The effect of the tangential component is
to induce a phase shift in the oscillation, leaving the amplitude
unchanged. The effect of the transversal component, sometimes
called \emph{oblique} component, is to modify the amplitude,
without affecting the phase of the oscillation.

Under the effect of the perturbation, the response of the
oscillator becomes
\begin{equation}\label{sec2-eq3}
\x(t) = \x_0(t + \psi(t)) + \z(t)
\end{equation}
where $\z(t)$ describes a small perturbation of the amplitude
which decays exponentially fast, while $\psi(t)$ is a phase shift
induced by the injected signal. It can be shown \cite{demir00}
that $\psi(t)$ is the solution of the nonlinear phase deviation
equation
\begin{equation}\label{sec2-eq3}
\dot{\psi}(t) = \vet_1^T(t + \psi(t)) \, \g(\x_0(t+\psi(t)),t)
\end{equation}
The vector $\vet_1(t)$ is the unique non trivial $T$--periodic
solution of the adjoint problem
\begin{equation}\label{sec2-eq4}
\dot \y (t) = -\D \f(\x_0(t))^T \, \y(t)
\end{equation}
satisfying the normalization condition
\begin{equation}\label{sec2-eq5}
\y^T(t) \, \f(\x_0(t)) = 1  \qquad \forall \, t\in \R^+.
\end{equation}

Noisy signals are modelled by the perturbation
\begin{equation}\label{sec2-eq6}
\g(\x(t),t) = \G(\x(t)) \, \Gamma(t)
\end{equation}
where $\G : \R^n \mapsto \R^{n\times m}$ is a state dependent
matrix and $\Gamma: \R \mapsto \R^m $ is the vector of noise
components. The corresponding phase deviation equation
\eqref{sec2-eq3} is a stochastic differential equation that
requires a probabilistic treatment. The time evolution for the
probability density function $p(\psi,t)$ is given by the
Fokker--Planck equation \cite{demir00}
\begin{eqnarray}\nonumber
\frac{\partial p(\psi,t)}{\partial t} & = &  -
\frac{\partial}{\partial \psi} \, \left( \vet(t+\psi)
\frac{\partial \vet(t+\psi)^T}{\partial \psi} \, p(\psi,t)
\right) \\[1ex]\label{sec2-eq7}
& & + \frac{\partial^2}{\partial \psi^2} \left( \vet(t + \psi)^T
\, \vet(t+\psi) \, p(\psi,t)\right) \qquad
\end{eqnarray}

\noindent where $\vet(t)^T = \vet_1(t)^T \G(\x(t))$.

Both eq. \eqref{sec2-eq4} and \eqref{sec2-eq7} show that an
analytical expression for the vector $\vet_1(t)$ is necessary to
write down the phase deviation equation and the Fokker--Planck
equation in a closed form. Unfortunately, in most of situations
this vector can only be determined numerically. In the next
sections we shall derive an analytical formula for the vector
$\vet_1$ in terms of the unperturbed vector field $\f(\cdot)$ and
the trajectory $\x_0(t)$ which holds for any planar oscillator.

\section{Diliberto's theorem}\label{diliberto theorem}

The Diliberto's theorem is a classical result on the integration
of planar homogeneous linear differential equations in terms of
geometrical quantities along a given trajectory of the system. For
a given vector field $\f = (f_1,f_2)^T$, we introduce the
perpendicular vector field $\f^{\perp} = (f_2,-f_1)^T$, and we
denote the jacobian matrix by $\A(t) = \D\f(\x_0(t))$.\\

\begin{theorem}[Diliberto's theorem \cite{chicone92}]
Let $\f(\x_0(t))$ be a solution of the variational equation
\begin{equation}\label{sec3-eq1}
\dot{\x}(t) = \A(t)\, \x(t)
\end{equation}
Then the fundamental matrix solution has the form
\begin{equation}\label{sec3-eq2}
\widetilde \X(t) = \left[\f(\x_0(t)) \quad a(t)
\f(\x_0(t))+\frac{b(t)}{||\f(\x_0(t))||^2} \, \f^{\perp}(\x_0(t))
\right]
\end{equation}
where $|| \f(\cdot) ||$ denotes the usual $L_2$ norm, $a(t)$ and
$b(t)$ are given by
\begin{eqnarray}\label{sec3-eq3}
b(t) & = & e^{\int_0^t \nabla \cdot \f(\x_0(s)) d s}\\[1ex] \nonumber
a(t) & = & \int_0^t \frac{\f^T(\x_0(s)) (\A(s) + \A(s)^T)
\f^{\perp}(\x_0(s))}{||\f(\x_0(s))||^4} \; b(s) \, ds
\\\label{sec3-eq4}
\end{eqnarray}

\noindent and $\nabla \cdot$ is the divergence operator.
\end{theorem}
See \cite{chicone92} for a proof of the theorem.

The Diliberto's theorem has important applications in the analysis
of planar nonlinear oscillations. If $\A(t) = \D\f(\x_0(t))$ is
the jacobian matrix of system \eqref{sec2-eq1} in absence of
perturbation, then eq. \eqref{sec3-eq1} determines the stability
of the limit cycle. It is trivial to verify that if $\x_0(t)$ is a
solution of eq. \eqref{sec2-eq1} for $\varepsilon = 0$, then
$\f(\x_0(t))$ solves eq. \eqref{sec3-eq1}. Thus one can make use
of Diliberto's theorem to find the fundamental matrix solution of
the variational equation \eqref{sec3-eq1} and to compute the
Floquet's multipliers, which determine the stability of the limit
cycle.
%

At each point $\x \in \bgamma$, the unperturbed oscillator's
(described by $\dot \x(t) = \f(\x(t))$) stable manifold can be
decomposed into two complementary linear spaces, the space $T
M_{\x}$ tangent to the unperturbed limit cycle at $\x$, and the
space $T I_{\x}$ tangent to the \emph{isochron} at $\x$
\cite{djurhuus09}. Let $\x_i(t)$, $i=0,1,\ldots$ be trajectories
(solutions) of the unperturbed system with initial condition
$\x_i$, with $\x_0 \in \bgamma$. The isochron based at $\x_0$ is
defined as the set of all initial conditions $\x_i$ such that the
trajectories $\x_i(t)$ asymptotically converge to $\x_0(t)$ on the
limit cycle (see figure \ref{figure1}). Extending the definition
to all points $\x \in \bgamma$ leads to the \emph{tangent bundles}
$T M$ and $T I$.
\begin{figure}[htb!]
\begin{center}
\includegraphics[width=40mm]{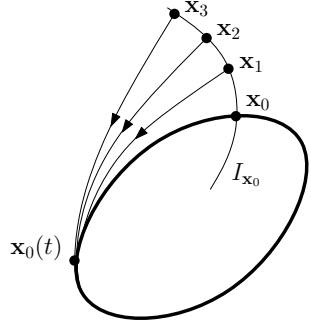}
\caption{Definition of an isochron.}\label{figure1}
\end{center}
\end{figure}
The tangent bundles $T M$ and $T I$ are spanned by the vectors
$\uvet_1(t)$ and $\uvet_2(t)$, respectively, which are the
Floquet's eigenvectors of the variational equation
\eqref{sec3-eq1} \cite{demir00,djurhuus09}\footnote{For a
$n$--order oscillator, $TI_{\x}$ is spanned by
$\{\uvet_2(t),\dots,\uvet_n(t)\}$.}. In general finding the
Floquet's eigenvectors requires several numerical integrations
\cite{demir00}, but in the planar case, we shall show that using
Diliberto's theorem analytical formulas for $\uvet_1(t)$ and
$\uvet_2(t)$ can be derived\footnote{The application to higher
order system is nontrivial because a full generalization of
Diliberto's theorem to higher dimensions is still an open issue.}.

\section{Phase space decomposition using Diliberto's
theorem}\label{basis}

In this section we derive analytical formulas, in terms of the
unperturbed limit cycle, for the vectors spanning the tangent
bundles $TM$ and $TI$.\\

\begin{theorem}
The Floquet's eigenvectors $\uvet_1(t)$ and $\uvet_2(t)$, spanning
the tangent bundles  $T M$ and $T I$ are given by
\begin{eqnarray}\label{sec4-eq1}
\uvet_1(t) & = & \f(\x_0(t))\\[1ex] \label{sec4-eq2}
\uvet_2(t) & = & e^{-\mu_2\, t} \left(\alpha(t) \, \f(\x_0(t)) +
\beta(t) \,  \f^{\perp}(\x_0(t))\right) \qquad
\end{eqnarray}

\noindent where $\mu_2$ is the second Floquet's exponent and
\begin{eqnarray}\label{sec4-eq2bis}
\alpha(t) & = & \frac{a(T)}{b(T) -1} + a(t) \\\label{sec4-eq2ter}
\beta(t) & = & \frac{b(t)}{||\f(\x_0(t))||^2}
\end{eqnarray}
with $a(T)$ and $b(T)$ given by \eqref{sec3-eq3} and
\eqref{sec3-eq4}.
\end{theorem}

\emph{Proof:}
for shorthand of notation, we introduce the vector field $\F : \R
\mapsto \R^n$ defined by $\F = \f \circ \x_0$, e.g. $\F(t) =
\f(\x_0(t))$.

It is straightforward to obtain the inverse of \eqref{sec3-eq2}
\begin{equation}\label{sec4-eq3}
\widetilde \X^{-1}(t) = \frac{1}{b(t)} \left[
\begin{array}{c}
-a(t) \, \F^{\perp}(t)^T +
\frac{b(t)}{||\F(t)||^2} \, \F(t)^T \\[2ex]
\F^{\perp}(t)^T
\end{array} \right]
\end{equation}

\noindent We can use \eqref{sec4-eq3} to construct the \emph{state
transition matrix} $\widetilde \X(t,0) = \widetilde \X^{-1}(0) \,
\widetilde \X(t)$, i.e. the fundamental matrix satisfying
$\widetilde \X(0) = \id$, where $\id$ is the identity matrix,
\begin{equation} \label{sec4-eq4}
\begin{array}{l}
\widetilde \X(t,0) = \\[1ex]
\left[
\begin{array}{cc}
\frac{\F(0)^T}{||\F(0)||^2}\, \F(t) &
\frac{\F(0)^T}{||\F(0)||^2}\, \left(a(t) \, \F(t) + \frac{b(t)}
{||\F(t)||^2} \, \F^{\perp}(t) \right) \\[2ex]
\F^{\perp}(0) \F(t) & \F^{\perp}(0)^T  \left( a(t) \, \F(t) +
\frac{b(t)}{||\F(t)||^2}\, \F^{\perp}(t)\right)
\end{array} \right]
\end{array}
\end{equation}

\noindent Keeping in mind that $\F(T) = \F(0)$ since $\f(\x_0(T))
= \f(\x_0(0))$, we derive the monodromy matrix
\begin{equation}\label{sec4-eq5}
\widetilde \X(T,0) = \left[
\begin{array}{cc}
1 & a(T) \\
0 & b(T)
\end{array} \right]
\end{equation}
The eigenvalues of the monodromy matrix are the characteristic
(Floquet's) multipliers $\lambda_i$, $i=1,2$. They are related to
the characteristic exponents $\mu_i$ by the relation
\begin{equation}\label{sec4-eq6}
\lambda_i = e^{\mu_i \, T}
\end{equation}
In this case, the multipliers are real and the exponents
are\footnote{For $n$--order systems, the Floquet's multipliers
are, in general, complex numbers. Therefore one has $\mu_i =
\frac{1}{T} \ln\left(|\lambda_i| + i (\arg \lambda_i + 2 k \pi)
\right)$ and an infinite number of characteristic exponents
corresponds to the same multiplier.}
\begin{equation}\label{sec4-eq7}
\mu_1 = 0 \hspace{20mm} \mu_2 = \frac{1}{T} \int_0^T \nabla \cdot
\f(\x_0(s)) \, ds
\end{equation}
where eq. \eqref{sec3-eq3} has been used.

According to Floquet's theory \cite{demir00}, the state transition
matrix can be written in the form
\begin{equation}\label{sec4-eq8}
\widetilde \X(t,0) = \widetilde \U(t) \, \exp (\D \, t) \,
\widetilde \U(0)^{-1}
\end{equation}
where $\widetilde \U(t)$ is a $T$--periodic matrix $\widetilde
\U(t) = \widetilde \U(t+T)$, such that $\widetilde \U(0)$ is the
matrix of the eigenvectors of the monodromy matrix $\widetilde
\X(T,0)$, i.e.
\begin{equation}
\mathbf{\Lambda} = \widetilde \U(0)^{-1} \, \X(T,0) \, \widetilde
\U(0)
\end{equation}
with $\mathbf{\Lambda} = \diag[\lambda_1,\lambda_2]$, and $\D =
\diag[\mu_1,\mu_2]$.

Computing the eigenvectors of \eqref{sec4-eq5} and from
\eqref{sec4-eq8} we derive
\begin{equation}\label{sec4-eq9}
\begin{array}{l}
\widetilde \U(t) = \left[
\begin{array}{cc}
\frac{\F(0)^T}{||\F(0)||^2}\, \F(t) & \frac{\F(0)^T}{||\F(0)||^2}
\left( \alpha(t) \, \F(t) + \beta(t) \, \F^{\perp}(t) \right)\\[2ex]
\F^{\perp}(0)^T \F(t) & \F^{\perp}(0)^T \left( \alpha(t) \, \F(t)
+ \beta(t) \, \F^{\perp}(t) \right)
\end{array}\right] \\[4ex]
\hspace{20mm} \times \exp(-\D\,t)
\end{array}
\end{equation}

\noindent where
\begin{eqnarray}\label{sec4-eq10}
\alpha(t) & = & \frac{a(T)}{b(T) -1} + a(t) \\\label{sec4-eq11}
\beta(t) & = & \frac{b(t)}{||\F(t)||^2}
\end{eqnarray}

It is well known that the state transition matrix is not unique,
because it can be constructed starting from different fundamental
matrices, corresponding to different initial conditions. However,
all the state transition matrices are similar, that is, if
$\X(t,0)$ and $\widetilde{\X}(t,0)$ are state transition matrices,
then a matrix $\C$ exists such that $\widetilde \X(t,0) = \C \,
\X(t,0) \, \C^{-1}$. On the one hand we have
\begin{equation}\label{sec4-eq12}
\X(t,0) = \U(t) \exp(\D \, t) \U(0)^{-1}
\end{equation}
and on the other hand
\begin{equation}\label{sec4-eq13}
\X(t,0) = \C^{-1} \, \widetilde \U(t) \exp(\D \, t) \, \widetilde
\U(0)^{-1} \, \C
\end{equation}
By comparison we have $\widetilde \U(t) = \C \, \U(t)$, and by
looking at \eqref{sec4-eq9} we can choose
\begin{equation}\label{sec4-eq14}
\C = \left[
\begin{array}{c}
 \frac{\F(0)^T}{||\F(0||^2} \\[1ex]
 \F^{\perp}(0)^T
 \end{array} \right]
 \end{equation}

\noindent The matrix $\U(t) = [\uvet_1(t),\, \uvet_2(t)]$ is given
by
\begin{equation}\label{sec4-eq15}
\U(t) = \left[ \F(t) \quad \alpha(t) \, \F(t) + \beta(t) \,
\F^{\perp}(t) \right] \exp(-\D \, t)
\end{equation}
and the theorem is proved.
\vspace{2mm}

Next we verify that $\U(t)$ is indeed $T$--periodic.\\

\begin{corollary}
The matrix $\U(t)$ is $T$--periodic
\end{corollary}

\emph{Proof:} it is obvious that the first column is periodic with
period $T$. The second column is periodic if and only if
\begin{equation}\left\{
\begin{array}{rcl}
\alpha(t + T) & = & \alpha(t) \, e^{\mu_2 \, T} \\[2ex]
\beta(t + T) & = & \beta(t) \, e^{\mu_2 \, T}
\end{array} \right.
\end{equation}

\noindent The second condition implies $b(t+ T) = b(t) e^{\mu_2 T}
= b(t) b(T)$, which is easily verified by using definition
\eqref{sec3-eq3}, the additivity of integrals and the fact that
$\nabla \cdot \f(\x_0(t))$ is a function of a periodic argument
and then it is also periodic.

The first condition leads to $a(t)\,b(T) = a(t+T) - a(T)$. Let us
denote
\begin{equation}
c(s) = \frac{\f(\x_0(s)) (\A(s) + \A(s)^T)
\f^{\perp}(\x_0(s))}{||\f(\x_0(s))||^4},
\end{equation}
obviously $c(s) = c(s+T)$. Using the definition of $a(t)$ given by
\eqref{sec3-eq4} and the additivity of integrals, the following
equality must hold
\begin{equation}
\int_0^t c(s) \, b(s) \, b(T) \, ds = \int_T^{T+t} c(r)\, b(r) \,
dr
\end{equation}
With the substitution $q=r-T$, using the periodicity of $c(s)$ and
the property $b(t+T) = b(t) b(T)$ we obtain an identity.

At each time instant the vectors $\uvet_1(t)$ and $\uvet_2(t)$ are
tangent to the unperturbed limit cycle and to the isochron,
respectively, at the point $\x_0(t)$, thus identifying the
tangential and the transversal direction into which the
perturbation must be decomposed.

\section{Determination of the reciprocal basis and the phase
deviation equation}\label{reciprocal basis}

Together with the set $\{\uvet_1(t),\ldots,\uvet_n(t)\}$ spanning
the tangent bundles $TM$ and $TI$, it comes another set
$\{\vet_1(t),\ldots,\vet_n(t)\}$, whose elements are defined as
the rows of the matrix $\V(t) = \U^{-1}(t)$. Thus by definition,
the bi--orthogonality condition
\begin{equation}\label{sec5-eq1}
\vet_i^T(t) \, \uvet_j(t) = \delta_{ij}
\end{equation}
holds. It follows that since $\{\uvet_1(t),\ldots,\uvet_n(t)\}$ is
a basis for $\R^n$, $\{\vet_1(t),\ldots,\vet_n(t)\}$ is a
\emph{reciprocal basis}, and its elements should be referred to as
\emph{covectors}. The covector $\vet_1(t)$ is normal to all
vectors $\uvet_k(t)$, $k=2,\ldots,n$ and in turn, each covector
$\vet_k(t)$, $k=2\ldots,n$ is normal to $\uvet_1(t)$. Thus
$\vet_1(t)$ spans the one--dimensional \emph{cotangent bundle}
$NI$ orthogonal to $TI$, while the covectors
$\{\vet_2(t),\ldots,\vet_n(t)\}$ span the $(n-1)$--dimensional
cotangent bundle $NM$ orthogonal to $TM$.\\

\begin{figure}[htb!]
\begin{center}
\includegraphics[bb=15 550 564 770, clip=true,width=80mm]{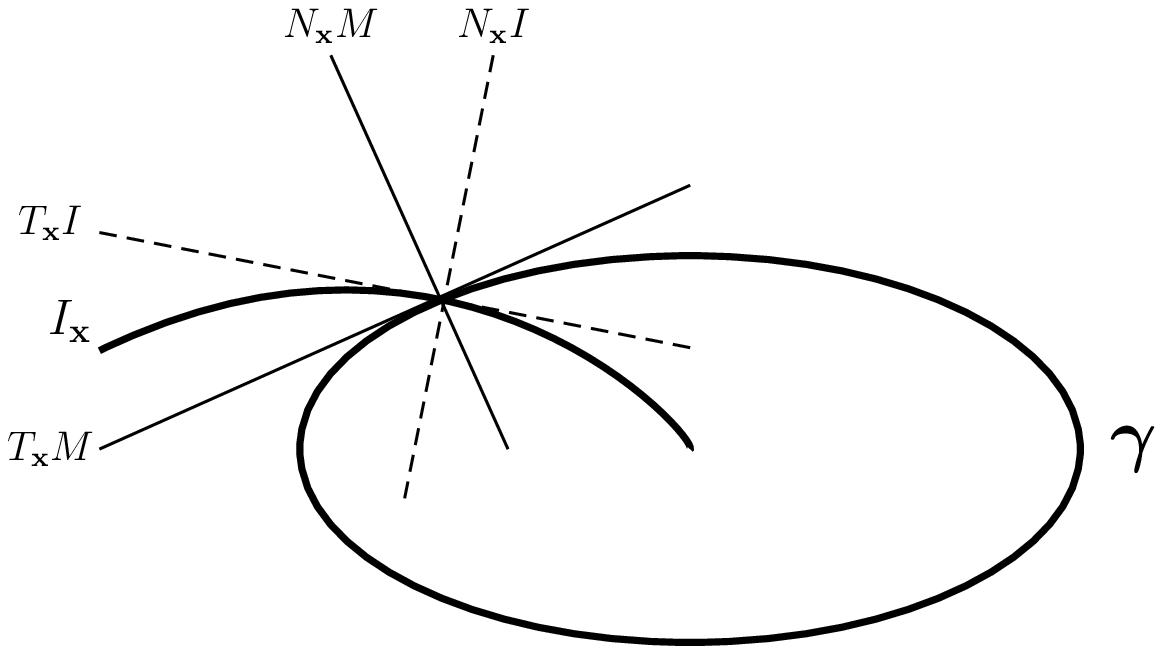}
\caption{Phase space decomposition of planar nonlinear
oscillations. The tangent bundles $T M$, $T I$ and the cotangent
bundles $N M$, $N I$ at the point $\x \in \bgamma$ are shown.
}\label{figure1}
\end{center}
\end{figure}

\begin{theorem}
The covector $\vet_1(t)$ entering in the phase deviation equation
\eqref{sec2-eq3} and the covector $\vet_2(t)$ are given by
\begin{eqnarray}\label{sec5-eq2}
\vet_1(t) & = & \frac{1}{b(t)} \left(-\alpha(t) \,
\f^{\perp}(\x_0(t)) + \beta(t) \, \f(\x_0(t)) \right)\\
\vet_2(t) & = & \frac{e^{\mu_2 \, t}}{b(t)} \,
\f^{\perp}(\x_0(t))\label{sec5-eq3}
\end{eqnarray}
\end{theorem}

\emph{Proof:}the inverse of \eqref{sec4-eq15} is
\begin{equation}
\V(t) = \frac{e^{\D\,t}}{\beta(t) ||\F(t)||^2} \left[
\begin{array}{c}
-\alpha(t) \F^{\perp}(t)^T + \beta(t) \F(t)^T\\[2ex]
\F^{\perp}(t)^T
\end{array} \right]
\end{equation}

\noindent from which \eqref{sec5-eq2} and \eqref{sec5-eq3} stem.
We shall now verify that $\vet_1(t)$ is the unique nontrivial
solution of the adjoint problem \eqref{sec2-eq4} satisfying the
normalization condition \eqref{sec2-eq5}. From \eqref{sec4-eq12}
we have $\V(t) = \exp(\D \, t) \, \V(0) \, \X(t,0)^{-1}$. Let us
introduce
\begin{equation}\label{sec5-eq3bis}
\Y(t,0) = \V(t)^T \exp(-\D\,t) \, \U(0)^T
\end{equation}
It is easy to verify that $\Y(t,0)^T \, \X(t,0) = \id$. This
implies that $\Y(t,0)$ is a state transition matrix of the adjoint
problem. We have
\begin{equation}\label{sec5-eq4}
\V(t)^T = \Y(t,0) \, \V(0) \, \exp(-\D\,t)
\end{equation}
which implies $\vet_1(t) = \Y(t,0) \vet_1(0)$. By taking the
derivative we get
\begin{equation}\label{sec5-eq4}
\dot{\vet}_1(t) = \dot{\Y}(t,0) \vet_1(0) = - \A^T(t) \Y(t,0)
\vet_1(0) = - \A^T(t) \vet_1(t)
\end{equation}
that is, $\vet_1(t)$ is a solution of the adjoint problem.
Finally, by definition $\V(t) \U(t) = \id$, that is $\vet_i(t)^T
\uvet_j(t) = \delta_{ij}$, which implies $\vet_1(t)^T \F(t) = 1$
as required.
\vspace{2mm}

The bi--orthogonality condition \eqref{sec5-eq1} implies that at
each time instant the covector $\vet_1(t)$ is parallel to the
gradient of the isochron. Thus $\vet_1(t)$ identifies the
direction along which the isochron is most sensitive to the
perturbations. At glance, eq. \eqref{sec2-eq3} and
\eqref{sec5-eq2} may look rather difficult to use in practical
application. However, one can take advantage of the periodicity of
$\vet_1(t)$ and consider its Fourier series
\cite{maffezzoni09,maffezzoni10}
\begin{equation}
\vet_1(t) = \sum_{k=-\infty}^{+\infty} V_k \, e^{i \,k \, \omega
\, t}
\end{equation}
In this case, eq. \eqref{sec5-eq2} can be a practical tool to
compute the spectral coefficients $V_k$.

\begin{figure}[htb!]
\begin{center}
\includegraphics[width=80mm]{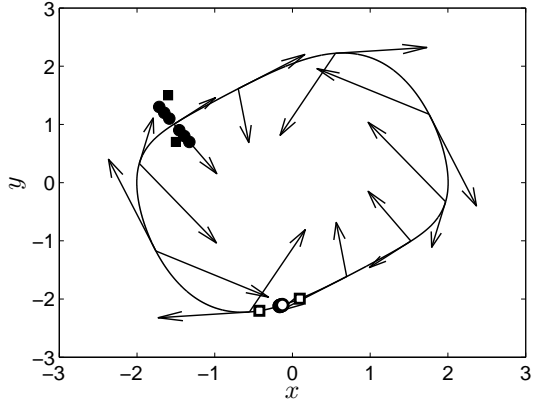}
\caption{Application of the method to the van der Pol oscillator.
The explanation is given in the text.}\label{figure3}
\end{center}
\end{figure}

Figure \ref{figure3} shows the application of the method to the
van der Pol oscillator. The vectors tangent and those transversal
to the cycle represent the vectors $u_1$ and $u_2$ respectively,
at different time instants. The filled circles represent the
initial conditions of trajectories starting on the plane spanned
by $u_2(\bar{t})$, and thus tangent to the isochron based at
$\bgamma(\bar t)$. The empty circles represent the same
trajectories after a long enough transient. All trajectories are
very close to the limit cycle and share the same phase. The filled
and the empty squares are the analogues for trajectories with
initial condition outside the isochron. In this case the final
points have different phases.

\section{On the validity of an orthogonal decomposition}\label{orthogonal decomposition}

From eqs. \eqref{sec4-eq1} and \eqref{sec4-eq2} we recognize that,
in general, $T M$ and $T I$ are not orthogonal spaces. This
explains why the original conjecture proposed in \cite{hajimiri98}
to decompose the perturbation into orthogonal components is, in
general, incorrect. However, it turns out that for a class of
oscillator the orthogonal decomposition is
appropriate.\\

\begin{theorem}
For all those oscillators such that $\f^T(\x_0(t)) \left(\A(t) +
\A^T(t) \right) \f^{\perp}(\x_0(t)) = 0$, the perturbation can be
decomposed into two orthogonal components.
\end{theorem}

\emph{Proof:} we have
\begin{equation}
\uvet_1(t)^T \uvet_2(t) = \alpha(t) \, ||\F(t)||^2 \, e^{\mu_2 t}
\end{equation}
which can be identically null if and only if $\alpha(t) = 0$ for
all $t \in \R$. By eq. \eqref{sec4-eq2bis} this implies $a(t) =
a(T)/(1-b(T))$, that is, $a(t)$ must be a constant. Since
$\X(T,0)$ is regular, we have $b(T) \ne 0$, and it stems that
$a(t)$ must be null for all $t$. This condition is obviously
satisfied if and only if $\F(t)^T \left(\A(t) +
\A^T(t)\right)\F^{\perp}(t) = 0$.
\vspace{2mm}

We shall now show that the condition above is equivalent to a
classical result about the admissibility of an orthogonal
decomposition. Some preliminary considerations are needing. The
isochrons represent the leaves of an invariant foliation of the
oscillator stable manifold \cite{djurhuus09}. A classical result
in nonlinear control theory states that if a given vector field is
tangent to a foliation, and another vector filed preserves this
foliation, then the Lie bracket of the vector fields is tangent to
the foliation \cite{jakubczyck01}. It stems that a necessary and
sufficient condition for $\f^{\perp}$ to generate an invariant
foliation of the oscillator's stable manifold is
\begin{equation}
[\f,\f^{\perp}](\x_0(t)) = c \, \f^{\perp}(\x_0(t))
\end{equation}
where $[\cdot,\cdot]$ represents the Lie bracket and $c \in \R$.\\

\begin{theorem}
The condition
\begin{equation*}
\f(\x_0(t)) \left(\A(t) + \A^T(t) \right) \f^{\perp}(\x_0(t)) = 0
\end{equation*}
is equivalent to $[\f,\f^{\perp}](\x_0(t)) = c \,
\f^{\perp}(\x_0(t))$
\end{theorem}

\emph{Proof:} we shall prove that the first condition implies the second.
The converse can be proved in the same way. For the sake of
simplicity we simply write $\f$, $\f^{\perp}$ and $\A$ omitting
the arguments $\x_0(t)$ and $t$. By hypothesis $(\A + \A^T)
\f^{\perp} = c^* \, \f^{\perp}$, for some $c^* \in \R$. We recall
that the Lie bracket of vector fields is defined by $[\f,\g] = \D
\g \, \f - \D \f \, \g$ where $\D \f$ and $\D \g$ are Jacobian
matrices. A routine calculation shows that
\begin{equation*}
[\f,\f^{\perp}] = \left((\nabla \cdot \f) \f^{\perp} -(\A + \A^T)
\f^{\perp} \right) = \left((\nabla \cdot \f) - c^* \right)
\f^{\perp}
\end{equation*}
and by setting $\left(\nabla \cdot \f - c^* \right) = c$ the proof
is completed.
\vspace{2mm}

\section{Conclusions}\label{conclusions}

In this brief we have presented a rigorous decomposition for the
phase space of weakly perturbed planar nonlinear oscillators. We
have derived analytical formulas for the basis spanning the
directions onto which project the perturbations. We have also
derived analytical formulas for the complementary basis formed by
the associated covectors. These equations have been derived using
a classical result on the integration of planar homogenous
differential equations, the Diliberto's theorem, they are rigorous
and expressed in terms of the unperturbed limit cycle, only. Since
they are exact, they can be used as benchmarks to test the
accuracy of numerical methods. Our results allow to write the
phase deviation equation and the Fokker--Planck equation derived
in \cite{demir00} in a closed analytical form. These equations can
be exploited to investigate the synchronization of a nonlinear
oscillator with external signals and the influence of noise on the
oscillator's output spectrum. This will be the topic of future
works.

\bibliography{IEEEabrv,phase&diliberto_biblio}
\bibliographystyle{ieeetran}

\end{document}